\documentclass[dviwindo]{PoS}

\usepackage{amsmath,amssymb}
\usepackage{graphicx}

\title{Dark Energy and Dark Matter in Stars Physic}

\ShortTitle{Dark Energy and Dark Matter in Stars Physic}

\author{\speaker{Plamen Fiziev} \\
        Jount Institute of Nuclear Research, Dubna, Russian Federation\\
        E-mail: \email{fiziev@theor.jinr.ru}}

\abstract{We present the basic equations and relations
 for the relativistic static spherically symmetric  stars (SSSS)
 in the  model of minimal dilatonic  gravity  (MDG)
 which is {\em locally} equivalent to the f(R) theories of gravity and gives an alternative
 description of the effects of dark matter and dark energy.
 The results for the simplest form of the relativistic equation of state (EOS) of neutron matter are represented.
 Our approach overcomes the well-known difficulties of the physics of SSSS in the f(R) theories of gravity
 introducing two novel EOS for cosmological energy-pressure densities
 and dilaton energy-pressure densities, as well as  proper boundary conditions.}

\FullConference{Frontiers of Fundamental Physics 14\\
                 15-18 July 2014\\
                 Aix Marseille University (AMU) Saint-Charles Campus, Marseille, France}

\begin{document}

\section{Introduction}
One of the most important lessons from the spectacular development of cosmology
in the last fifteen years is the clear understanding that the Einstein general relativity (GR)
and standard particle model (SPM)
are insufficient to explain all observed phenomena in the Nature.
There exist three possible ways for further development:

1) To add some new content of the Universe beyond the SPM, like dark matter and dark energy;

2) To change the theory of gravity. The simplest models are $f(R)$ \cite{fR}  and MDG \cite{OHanlon72,Fiziev00a};

3) Some mixture of these two possibilities is not excluded by the current
observational data.

Dozens of models with different functions $f(R)$ exist, some of them dubbed valuable \cite{fR}.

The situation in star physics is similar.
More than sixty-year development so far has not solved the problem with realistic EOS
of compact star matter. At present one can find several dozens of EOS in the literature.

A series of attempts to use  $f(R)$ models of gravity adopted to star physics also exist
\cite{fRstars}. No fully convincing final result was reached.

The main goal of the present paper is to create a clear physical basis
for application of MDG in star physics and thus to
facilitate construction of models, which permit unified treatment
of the physical problems at very different scales:
from laboratory scales and compact star scales to the scale of the visible Universe.
Such a unified approach may give much more definite justification of our models
using all available information for the physical phenomena at all reachable scales.

The MDG model was proposed and studied in \cite{OHanlon72,Fiziev00a,MDG,Fiziev13}.
It describes a simple generalization of the Einstein general relativity (GR),
based on the following action of the gravi-dilaton sector
\begin{eqnarray}
{\cal A}_{g,\Phi}={\frac {c^3} {16\pi G_{N}}}\int d^4 x\sqrt{|g|}
 \bigl( \Phi R - 2 \Lambda U(\Phi) \bigr).
\label{A_MDG}
\end{eqnarray}
Without any relation with astrophysics and cosmology,
it was studied by O'Hanlon, as early as in \cite{OHanlon72}.
There the term "dilaton" for the field $\Phi$ was introduced.
Formally, MDG resembles the Branse-Dicke theory with $\omega\equiv 0$, if
the most important MDG-cosmological term in Eq. (\ref{A_MDG}) is ignored.
To some extent MDG is related with the $f(R)$
theories of gravity \cite{fR,fRstars}.
In general, the $f(R)$ theories are also physically different,
being only {\em locally} equivalent to MDG \cite{Fiziev13}.

The values $\Phi\in (0,\infty)$ must be positive to avoid
the physically unacceptable antigravity.
The value $\Phi=0$ yields an infinite gravitational factor and makes the Cauchy problem
in MDG not well posed \cite{EFarese}.
The value $\Phi=\infty$ turns off the gravity and must also be excluded, as well as $\Phi=0$.

The scalar field $\Phi$ introduces a variable gravitational factor $G(\Phi)=G_N/\Phi=G_N g(\Phi)$ instead
of the Newton constant $G_N$.
The cosmological potential $U(\Phi)$  introduces a variable factor $\Lambda U(\Phi)$
instead of the constant $\Lambda$. In GR with cosmological constant $\Lambda$
we have $\Phi\equiv 1$, $g(\Phi)\equiv 1$, and $U(1)\equiv 1$.
Due to its specific physical meaning,
the field $\Phi$ has quite unusual properties.

The function $U(\Phi)$ defines the cosmological potential
which must be a positive {\em single valued} function of the dilaton field $\Phi$ by astrophysical reasons.
See \cite{Fiziev13} for all physical requirements on the cosmological potential $U(\Phi)$,
necessary for a sound MDG model.
There the class of {\em withholding} potentials was introduced.
These confine dynamically the values of the dilaton $\Phi$ in the physical domain.
It is hard to formulate such a property for the function $f(R)$ in a simple way.

Some more physical and astrophysical consequences of MDG are described in
\cite{OHanlon72,Fiziev00a,MDG,Fiziev13,EFarese}.
This model provides an alternative explanation of the observed astrophysical phenomena
without introduction of dark energy and dark matter.

In the the present article we give a correct formulation of a star problem in MDG
and consider the simplest example of a physically consistent family of SSSS
for an admissible cosmological potential $U(\Phi)$.
We explicitly show how the dilatonic field $\Phi$ changes the structure of the compact stars
and creates a specific dilatonic sphere around them, analogous to dark matter halo.

\section{Basic equations and boundary conditions for SSSS in MDG}
In units $G_N=c=1$  the field equations of MDG can be written in the form:
\begin{subequations}\label{DGE:ab}
\begin{eqnarray}
\Phi \hat{R}_\alpha^\beta +\widehat{\nabla_\alpha\nabla^\beta}\Phi+ 8\pi\hat{T}_\alpha^\beta&=&0, \label{DGE:a}\\
\Box\Phi+ \Lambda V^\prime(\Phi) &=& {\frac {8\pi} 3} T. \label{DGE:b}
\end{eqnarray}
\end{subequations}
Here ${T}_\alpha^\beta$ is the energy-momentum tensor of the matter,
$\hat{X}_\alpha^\beta={X}_\alpha^\beta - {\frac 1 4}X\delta_\alpha^\beta$
is the traceless part of the $4D$-tensor ${X}_\alpha^\beta$,
$X=X_\alpha^\alpha$ is its trace,
the relation $V^\prime(\Phi)={\frac 2 3}\Big(\Phi U^\prime(\Phi) -2U(\Phi)\Big)$
introduces the dilatonic potential $V(\Phi)$, and
the prime denotes differentiation with respect to the variable $\Phi$ \cite{Fiziev13}.

In the problems under consideration, the space-time-interval is
$ds^2=e^{\nu(r)}dt^2-e^{\lambda(r)}dr^2 - r^2 d\Omega^2 $ \cite{Luciano},
where $r$ is the luminosity distance to the center of symmetry,
and $d\Omega^2$ describes the space-interval on the unite sphere.
Then, after some algebra one obtains  the following basic results
for a SSSS of the luminosity radius $r^*$.

In the inner domain   $r\in [0,r^*]$  the SSSS structure is determined by the system:
\begin{subequations}\label{DE:abcd}
\begin{eqnarray}
{\frac {dm}{dr}}&=&4\pi r^2\epsilon_{eff}/\Phi, \label{DE:a}\\
{\frac {d\Phi}{dr}}&=&-4\pi r^2 p_{{}_\Phi}/\Delta,  \label{DE:b}\\
{\frac {dp_{{}_\Phi}}{dr}}&=&- {\frac{ p_{{}_\Phi}}{r\Delta}}\left(3r -7 m-{\frac 2 3}\Lambda r^3+4\pi r^3\epsilon_{eff}/\Phi\right)
-{\frac{2}{r}}\epsilon_{{}_\Phi}, \label{DE:c}\\
{\frac {dp}{dr}}&=&- {\frac {p+\epsilon}{r}}\,{\frac{m+4\pi r^3 p_{eff}/\Phi}{\Delta-2\pi r^3 p_{{}_\Phi}/\Phi} }. \label{DE:d}
\end{eqnarray}
\end{subequations}
The four unknown functions are $m(r)$, $\Phi(r)$, $p_{{}_\Phi}(r)$, and $p(r)$. In Eqs. (\ref{DE:abcd})
$\Delta=r-2 m-{\frac 1 3}\Lambda r^3$,
$\epsilon_{eff}=\epsilon+\epsilon_{{}_\Lambda}+\epsilon_{{}_\Phi}$, $p_{eff}=p+p_{{}_\Lambda}+p_{{}_\Phi}$.
We obtain also two additional EOS specific for MDG:
\begin{subequations}\label{NewEOS:abc}
\begin{eqnarray}
\epsilon_{{}_\Lambda}&=&- p_{{}_\Lambda}-{\frac \Lambda {12\pi}}\Phi
\,\,\,: \hskip 5.5truecm\text{CEOS};\label{NewEOS:a}\\
\epsilon_{{}_\Phi}&=&p-{\frac 1 3}\epsilon +
{\frac \Lambda {8\pi}}V^\prime(\Phi)
+{\frac {p_{{}_\Phi}} 2}\,{\frac{m+4\pi r^3 p_{eff}/\Phi}{\Delta-2\pi r^3 p_{{}_\Phi}/\Phi}}
\,\,\,: \hskip 1.truecm\text{DEOS}; \label{NewEOS:b}\\
\epsilon\,\,\, &=& \epsilon(p)
\,\,\,: \hskip 6.9truecm\text{MEOS}.\label{NewEOS:c}
\end{eqnarray}
\end{subequations}

Equation (\ref{NewEOS:a}) is the EOS for the cosmological energy density
$\epsilon_{{}_\Lambda}={\frac \Lambda {8\pi}} \Big(U(\Phi)-\Phi\Big)$
and the cosmological pressure
$p_{{}_\Lambda}=-{\frac \Lambda {8\pi}} \Big(U(\Phi)-{\frac 1 3}\Phi\Big)$,
i.e.  the cosmological EOS (CEOS).

Equation (\ref{NewEOS:b}) replaces the field equation (\ref{DGE:b}) and presents the dilatonic EOS (DEOS).
It relates the dilatonic energy density
$\epsilon_{{}_\Phi}={\frac 1 {8\pi r^2}}\left(\Delta/r\right)^{1/2}{\frac d {dr}}\left(r^2 \left(\Delta/r\right)^{1/2}{\frac d {dr}}\Phi\right)$
and  the dilatonic pressure   $p_{{}_\Phi}=-{\frac \Delta {4\pi r^2}}  {\frac d {dr}}\Phi$, which
measures the gradient of gravitational factor.

Equation (\ref{NewEOS:c}) presents the EOS of star matter (MEOS), see \cite{Luciano} for a modern detailed survey.

Adopting the widespread assumption that the SSSS-center C ($\Rightarrow$ index "c") is at $r_c=0$ we obtain the boundary conditions
\begin{eqnarray}
m(0)=m_c=0,\,\,\,\,\,\Phi(0)=\Phi_c,\,\,\,\,\,p(0)=p_c,\nonumber \\
p_{{}_\Phi}(0)=p_{\Phi c}={\frac 2 3}\left({\frac {\epsilon(p_c)} 3}- p_c\right)-{\frac{\Lambda}{12\pi}}V^\prime(\Phi_c).
\label{center}
\end{eqnarray}
Requiring $m_c=0$ we ensure finiteness of pressure $p_c$ simultaneously for the Newton-, GR- and MDG-SSSS.
The condition on $p_{\Phi c}\,(=-{\frac 2 3}\epsilon_{\Phi c})$  ensures its finiteness,
being a specific MDG-centre-value-relation: $F_\Phi(p_{\Phi c},p_c,\Phi_c)=0$.

The SSSS-edge ($\Rightarrow$ index "*")  is defined by the condition $p^*=p(r^*;p_c,\Phi_c)=0$. Then
\begin{eqnarray}
m^*=m(r^*;p_c,\Phi_c),\,\,\,\Phi^*=\Phi(r^*;p_c,\Phi_c),\nonumber \\
p_{{}_\Phi}^*=p_{{}_\Phi}(r^*;p_c,\Phi_c).
\label{edge}
\end{eqnarray}
The luminosity radius of a compact SSSS varies: $r^*\sim 5\div 20$ km, depending on MEOS.

Outside the star $p\equiv 0$ and $\epsilon\equiv 0$, and we have a dilaton-sphere,
in brief -- {\em a disphere}. Its structure is determined
by the shortened system (\ref{DE:abcd}, i.e., the Eq. (\ref{DE:d}) has to be omitted.
For the exterior domain $r\in [r^*,r_{{}_{\!U}}]$ we use Eqs. (\ref{edge}) as left boundary conditions.
The right boundary is defined by the cosmological horizon $r_{{}_{\!U}}$:  $\Delta(r_{{}_{\!U}}; p_c,\Phi_c)=0$,
where the de Sitter vacuum is reached:
$\Phi(r_{{}_{\!U}}; p_c,\Phi_c)=1$.
Thus, we obtain a new MDG-centre-value-relation:   $F_\Lambda(p_c,\Phi_c)=0$.

Hence, in MDG, as well as in the Newton gravity and GR, we have a one-parameter-family of SSSS,
because of the existence of the two MEOS-dependent-relations:
\begin{eqnarray}
F_\Phi(p_{\Phi c},p_c,\Phi_c)=0,\,\,\,F_\Lambda(p_c,\Phi_c)=0.
\label{F}
\end{eqnarray}

The observable  value $\Lambda\approx 1.27\times 10^{-44}$ $km^{-2}$ is very small.
As a result, the luminosity radius of the Universe $r_{{}_{\!U}} \sim 1/\sqrt{\Lambda}\sim 10^{22}\,\text{km}$
is very large in comparison with $r^*\sim 10$ km.

Further on, we use the cosmological potential
$U(\Phi)=\Phi^2+{\frac{3}{16\, \mathfrak{p}^2}}\left(\Phi-1/ \Phi\right)^2$. Hence,
$V^\prime(\Phi)={\frac 1 {2 \mathfrak{p}^2}}\left(1-1/\Phi^2\right)$.
For useful comments and a more general form of the admissible cosmological potentials $U(\Phi)$ see \cite{MDG,Fiziev13}.
The parameter $\mathfrak{p}=\sqrt{\Lambda}\,\hbar/c\,m_\Phi$ is the dimensionless
Compton length (measured in cosmological units) of the dilaton $\Phi$.
MDG is consistent with observation if $\mathfrak{p}\lesssim 10^{-30}$ ($m_\Phi> 10^{-3}\, \text{eV}/c^2)$ \cite{Fiziev00a,MDG}.
If one takes as a dilaton the only currently known fundamental scalar particle,
the Higgs boson with mass $m_H \approx 125\, GeV/c^2$, then $\mathfrak{p} \approx 1.8\times 10^{-43}$.
Further on, we  use a non-realistic big value $\mathfrak{p} =  10^{-21}$ (the Compton length $\sim 9$ km) for a more transparent
graphical representation of the results.

\section{Results for the simplest MEOS of neutron matter}

\begin{figure}[!ht]
%\centering
\hskip 1.truecm
\begin{minipage}{5.cm}
\vskip .2truecm
\includegraphics[width=4.5truecm]{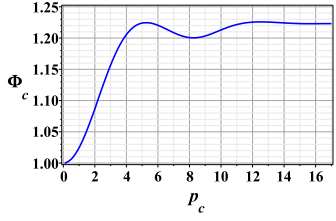}
\vskip -.2truecm
\caption{\small The specific MDG-curve \hskip .5truecm $F_\Lambda(p_c,\Phi_c)=0$}
\label{FLambda}
\end{minipage}
\end{figure}

\begin{figure}[!ht]
%\centering
\hskip 10.truecm
\begin{minipage}{5.cm}
\vskip -4.4truecm
\includegraphics[width=4.5truecm]{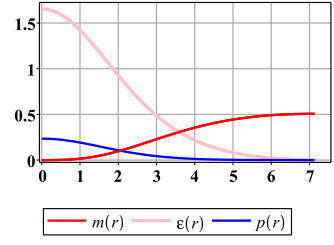}
\vskip -.3truecm
\caption{\small The MDG-SSSS interior in accord with MEOS}
\label{mep}
\end{minipage}
\end{figure}

\begin{figure}[!ht]
\vskip -.5truecm
\hskip 1.5truecm
\begin{minipage}{5.cm}
\includegraphics[width=4.5truecm]{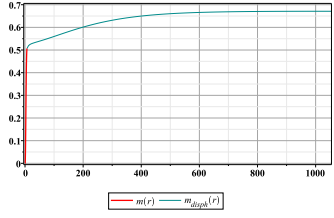}
\vskip .truecm
\caption{\small The disphere-mass dependence on $r$}
\label{mmdi}
\end{minipage}
\end{figure}

\begin{figure}[!ht]
\hskip 10.truecm
\begin{minipage}{5.cm}
\vskip -4.5truecm
\includegraphics[width=4.5truecm]{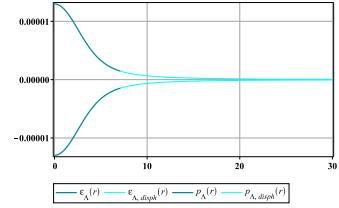}
\vskip .truecm
\caption{\small The dependencies $\epsilon_\Lambda(r)$ and $p_\Lambda(r)$ in accord with CEOS}
\label{eLpL_st_r_st}
\end{minipage}
\end{figure}

\begin{figure}[!ht]
\vskip -.5truecm
\hskip 1.5truecm
\begin{minipage}{5.cm}
\includegraphics[width=4.5truecm]{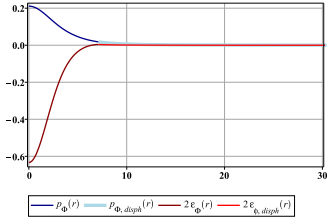}
\vskip .truecm
\caption{\small The dependencies $2\epsilon_\Phi(r)$ and $p_\Phi(r)$ in accord with DEOS}
\label{pPhiePhi}
\end{minipage}
\end{figure}

\begin{figure}[!ht]
\hskip 10.truecm
\begin{minipage}{6.cm}
\vskip -5.truecm
\includegraphics[width=4.5truecm]{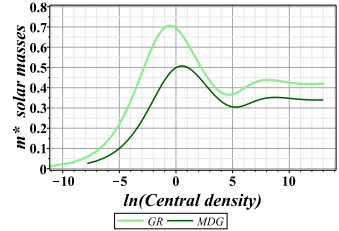}
\vskip .truecm
\caption{\small The m* -- $\epsilon_c$ dependencies}
\label{m centerD}
\end{minipage}
\end{figure}

\begin{figure}[!ht]
\vskip -.3truecm
\hskip 1.5truecm
\begin{minipage}{6.5cm}
\includegraphics[width=4.5truecm]{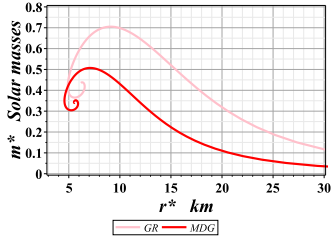}
\vskip .truecm
\caption{\small Mass - radius relations: in
 GR ($m_{max}^*\approx .7051\,m_{\odot}$, $r_{max}^*\approx 9.209$ km) and
in MDG ($m_{max}^*\approx .5073\,m_{\odot}$, $r_{max}^*\approx 7.092$ km)}
\label{m_r}
\end{minipage}
\vskip 1.truecm
\end{figure}

\begin{figure}[!ht]
\vskip -6.7truecm
\hskip 10.truecm
\begin{minipage}{5.cm}
\includegraphics[width=4.5truecm]{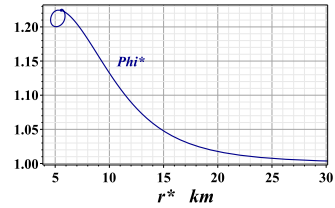}
\vskip .truecm
\caption{\small The specific MDG-dependence $\Phi^*(r*)$}
\label{Phi_st_r_st}
\end{minipage}
\vskip 1.truecm
\end{figure}

\begin{figure}[!ht]
\vskip -.truecm
\hskip .truecm
\begin{minipage}{8.2cm}
\hskip 2.truecm
\includegraphics[width=4.5truecm]{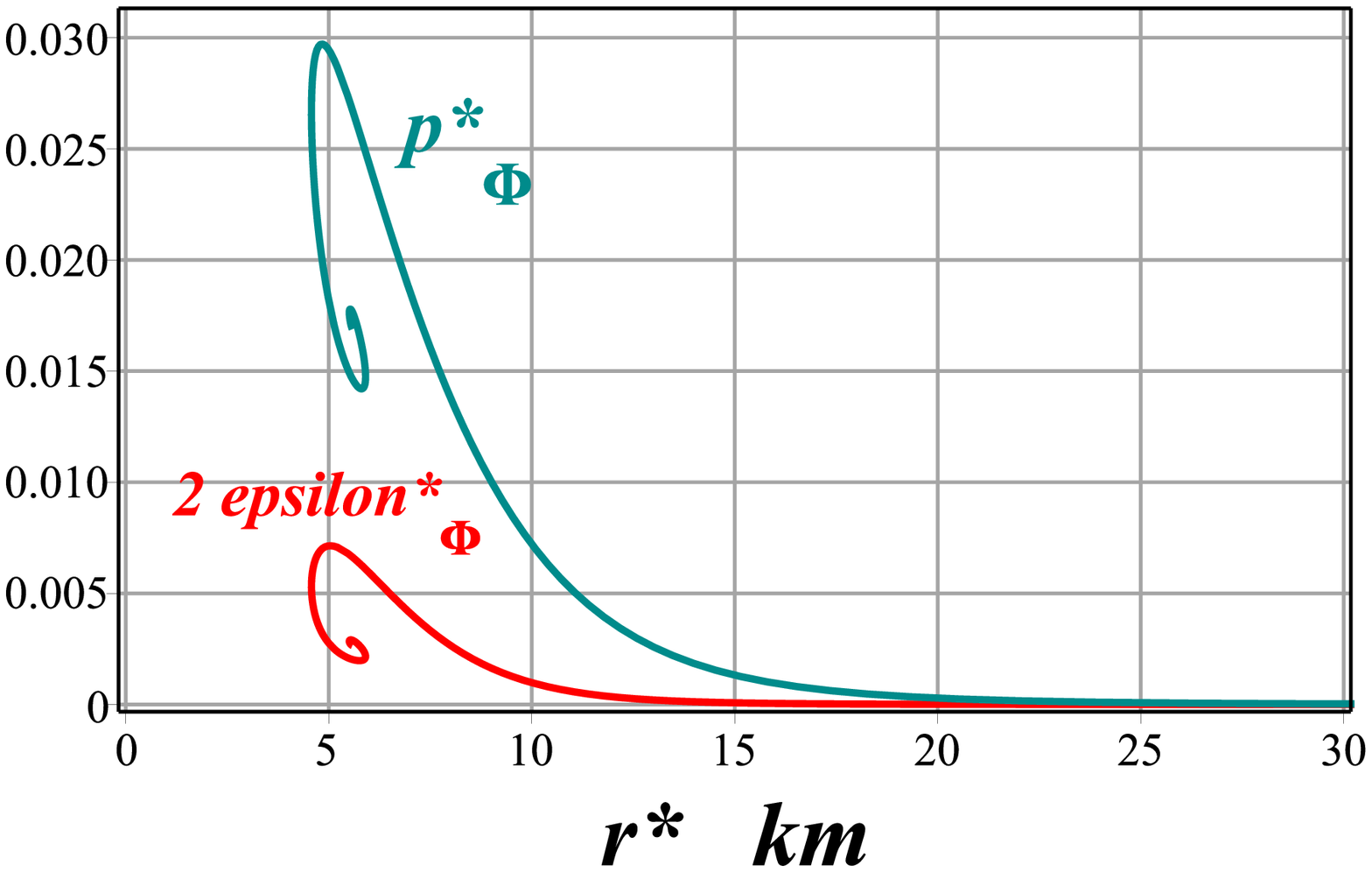}
\vskip .truecm
\caption \small The MDG-dependences $2 \epsilon_\Phi^*(r*)$, $p_\Phi^*(r*)$.
Maximal values are reached for smaller $r*$ than in Fig. \ref{m_r}
\label{epPhi_st_r_st}
\end{minipage}
\end{figure}

\begin{figure}[!ht]
\vskip -5.3truecm
\hskip 9.truecm
\begin{minipage}{6.5cm}
\hskip .7truecm
\includegraphics[width=4.9truecm]{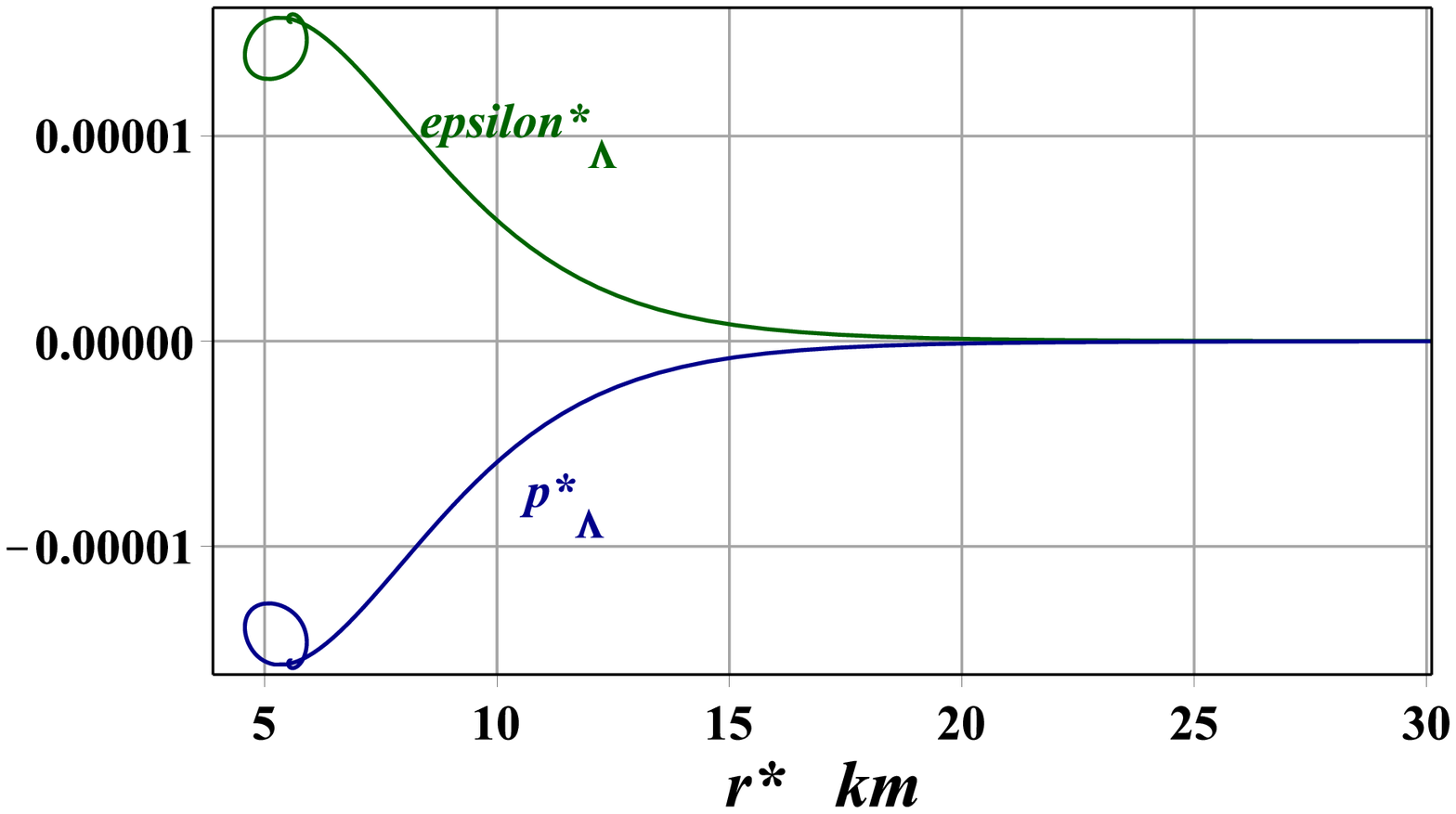}
\vskip .truecm
\caption \small The specific MDG-dependences $\epsilon_\Lambda^*(r*)$ and $p_\Lambda^*(r*)$ in accord with CEOS
\label{eLpL}
\end{minipage}
\end{figure}

\begin{figure}[!ht]
\vskip .3truecm
\hskip .truecm
\begin{minipage}{15.cm}
\hskip 4.5truecm
\includegraphics[width=5.truecm]{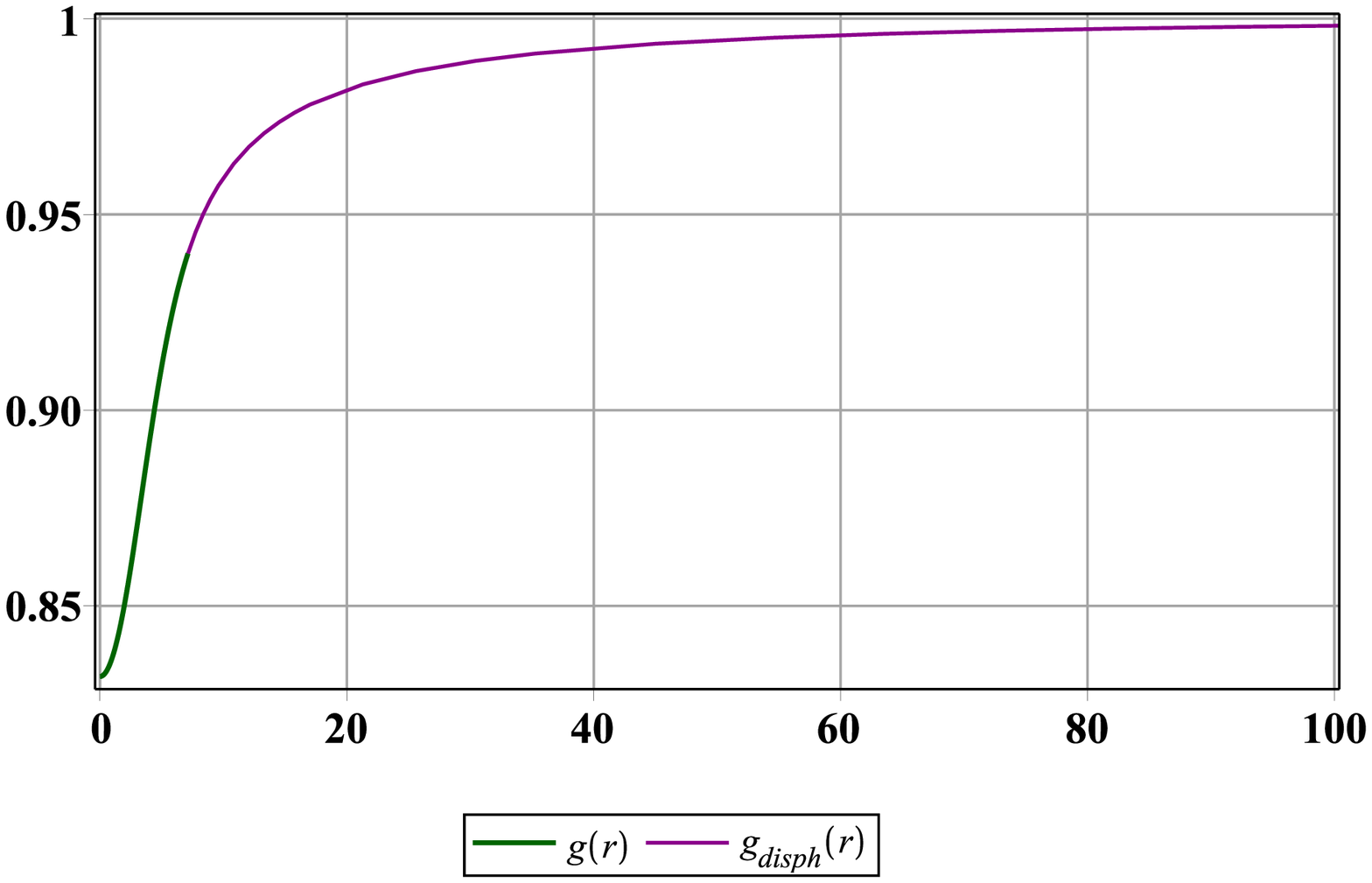}
\vskip .truecm
\caption{\small Unexpectedly large variations of the  gravitational
factor $g(r)=1/\Phi(r)$ in MDG-SSSS}
\label{gr}
\end{minipage}
\end{figure}

The most idealized MEOS of neutron matter is the one
from the GR-TOV model  \cite{Luciano}: $\epsilon={\frac 1 {4\pi}x\left( \sinh t -t\right)}$,
$p={\frac 1 {12\pi}\left( \sinh t -8\sinh(t/2)+3t\right)}$.
It describes the ideal Fermi neutron gas at zero temperature and
facilitates our study of the pure-MDG-effects in SSSS, shown in Figs. \ref{FLambda}--\ref{gr}.
%1
%

%2
%

%3
%
%4
%
%5
%
%6
%
%7

%8
%

%%%%%%%%%%%%%%%%%%%%%%%%%%%%%%%%%%%%%%%%%%%%%%%%%%%%%%%%%%%%%%
%9
\newpage
%10
%

%11

As seen in  Fig. \ref{m_r}, MDG-SSSS are lighter and more compact than GR-SSSS.
As in GR, MDG-SSSS may be stable only until maximal mass is reached, see Figs. \ref{m centerD} and \ref{m_r}.
The mass of the disphere $m_{disp}(r)$ outside MDG-SSSS
exponentially goes to a constant: $m_{disp} \approx .1638\, M_\odot$.
The total mass of the object $m_{total} \approx 0.671\, M_\odot$ is quite close to the mass of GR-SSSS $m_{GR,max} \approx 0.705\, M_\odot$.

An interesting problem is a model of a moving and rotating star in MDG.
In it one can expect an asymmetric configuration with
appearance of different centers of the star and its disphere,
or even detachment of the parts of disphere.
For similar effects at cosmological scales see \cite{Clowe}.

This research was supported in part by the Foundation for
Theoretical and Computational Physics and Astrophysics
and Grant of the Bulgarian Nuclear Regulatory, Agency for 2014,
as well as by "NewCompStar", COST Action MP1304.

\end{document}